\documentclass[aps,twocolumn,superscriptaddress]{revtex4}

\usepackage{amsmath,amsfonts,stmaryrd,amssymb,graphics,epsfig,txfonts}
\begin{document}

\title{Complete positivity and contextuality of quantum dynamics}

\author{Song Cheng}
\email{shandongchengsong@gmail.com}
\affiliation{School of Physics, Shandong University, Jinan 250100, China}
\author{Dong-Sheng Wang}
\email{wdscultan@gmail.com}
\affiliation{Institute for Quantum Science and Technology, \\Department of Physics and Astronomy, University of Calgary, Alberta T2N 1N4, Canada}

\begin{abstract}
Positivity or the stronger notion of complete positivity, and contextuality are central properties of quantum dynamics.
In this work, we demonstrate that a physical unitary-universe dilation model could be employed to characterize the completely positive map, regardless of the initial correlation condition.
Particularly, the problem of initial correlation can be resolved by a swap operation.
Furthermore, we discuss the physical essence of completely positive map and highlights its limitations.
Then we develop the quantum measurement-chain formula beyond the framework of completely positive map in order to describe much broader quantum dynamics, and therein the property of contextuality could be captured via measurement transfer matrix.
\end{abstract}

\date{20 March, 2013}

\maketitle

\section{Introduction}\label{sec:intr}

Quantum dynamics, which lies in Hilbert space, is explicitly distinct from classical mechanics and stochastic dynamics, although there are some similarities and connections among them.
As the properties of quantum dynamics, superposition and uncertainty have been explored greatly, while nowadays along with the development of quantum control and computation \cite{NC,Breuer} etc, the notions of complete positivity, entanglement, nonlocality, and also contextuality are of fundamental concern and more interest to researchers.

The generic frameworks to describe quantum dynamics include theories of quantum channel \cite{Stinespring,Choi} and quantum trajectory \cite{Chattaraj}, in this work we focus mainly on the former one.
As quantum state, namely density matrix, could be treated as one positive semi-definite operator, quantum dynamics is characterized as a map acting on the state.
A formal theory based on positive and completely positive (CP) maps provides a well-defined mathematical framework \cite{Stinespring,Choi}, from which the microscopic Markovian dynamics can be derived associated with a semigroup structure \cite{Lindblad,GKS,Alicki}.
The CP map, which is also known as a quantum channel, corresponds to the operator-sum representation (OSR) \cite{Kraus}.
However, the limitation is that CP map is not enough to describe all quantum dynamics, since CP map is only a subset of positive map.
Recently the limitations of CP map cause much attention, namely, the initial correlation issue.
It is proved that a map with initial classical-quantum state, which has null quantum discord, can be written as a CP map \cite{Sudarshan08,Lidar09}.
We will show that, via a swap operation trick, the open-system dynamics can always be written as a CP map when there is any initial correlation.

We analyze the physical properties of CP map employing a ``unitary-universe dilation model''.
We then show that the dilation model has limitations when applied to describe all quantum dynamics, for which we rather explore a different framework for the rest of the work.
It is proved that, beyond the dilation model, the OSR form with state-dependent Kraus operators could be constructed for a positive, non-CP map between any two quantum states \cite{tong04}.
The state-dependent OSR form could be understood as state-dependent quantum measurement with memories and feedbacks.
One constructive and operational quantum measurement-chain formula is developed in this paper to describe quantum dynamics beyond the dilation model.
The quantum measurement transfer matrix, which is the analogue of transfer matrix in classical stochastic dynamics, captures the contextuality of quantum dynamics.
Contextuality, originating from the Kochen-Specker theorem \cite{ks}, can be described via inequalities \cite{Cabello} or measures of contextuality \cite{measurec}.
A common example of contextuality of quantum observable is that for observable $A$ commuting with $B$ and $C$, which do not commute with each other, the measurable values of $A$ can be different when it is measured together with $B$ or $C$.
Technically, the reason is that one operator can have different decompositions, i.e. fine-grained structures, which can lead to physical effects in practice.
In this work we take a generalized notion of contextuality \cite{Spekkens}, without focusing on the exact distinctions of different notions.
Briefly, we say there exists contextuality of one operator (density matrix, observable, POVM (positive operator-valued measurement) etc), if the fine-grained structure of the operator causes measurable physical effects.

In this work, we study issues relating to properties of complete positivity and contextuality, by employing the unitary-universe model and quantum measurement-chain formula, respectively. The paper mainly contains two parts. In section \ref{sec:cpmap}, we study the mathematical framework for CP map, and particularly analyze the initial correlation problem, which can be solved via a swap trick in the unitary-universe dilation model. We define a physically positive map to address the limitations of CP map. Also, the limitations of Bloch vector dynamics in Heisenberg picture are discussed. In section \ref{sec:contex}, we study the quantum measurement-chain formula, and properties of measurement and weak-measurement transfer matrices.
The contextual property of quantum dynamics is demonstrated. At last we conclude briefly in section \ref{sec:conc}.

\begin{figure}
\centering
\includegraphics[scale=0.35]{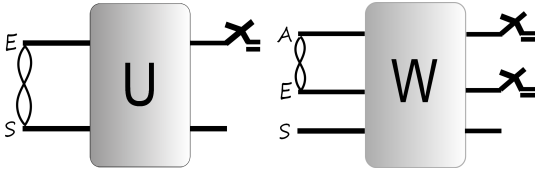}
\caption{Schematic diagram for the unitary-universe dilation model. $U$ and $W$ represent unitary operations. The meters represent trace operations. We introduce in the notation for convenience, that if two systems $S$ and $E$ are correlated, we denote $S\infty E$, and if two systems $S$ and $E$ interact (do not interact), we use $S\leftrightarrow E$ ($S\nleftrightarrow E$).}
\label{fig:uu}
\end{figure}

\section{Completely positive map}\label{sec:cpmap}

In this section, we briefly review the standard framework of CP map, and then establish the unitary-universe dilation model to interpret the physics.
For the initial correlation problem, we employ the swap operation and discuss one example to demonstrate its viability.
Then we discuss the properties of physically positive map and hermitian map, and highlight their limitations.

According to operator theory, quantum state $\rho$ is treated as a trace class, positive semidefinite operator acting on Hilbert space $H$, and $\rho\in \mathcal{T}(H)$. A map $\Phi:\mathcal{T}(H)\rightarrow\mathcal{T}(H)$ is positive iff it maps a state $\rho$ into a state $\Phi(\rho)$. A map $\Phi$ is completely positive (CP) iff $\Phi\otimes {\bf 1}_m$ is positive for any $m\in \mathbb{Z}^+$. Stinespring dilation theorem and Choi's theorem \cite{Stinespring,Choi,Bhatia} state that a CP map takes the form
\begin{align}
    \Phi(\rho)=\sum_{i=1}^N K_i \rho K_i^{\dagger},
\end{align}
and the dilation is said to be {\em minimal} for $N=n^2$, where $n$ is the dimension of the Hilbert space $H$. If $\sum_{i=1}^{N} K_i^{\dagger}K_i={\bf 1}$, the map is trace-preserving (TP). Note that a CP map $\Phi$ induces the CP map $\Phi\otimes{\bf 1}_m$, i.e. if map $\Phi$ is CP, then for all $m$, the positive map $\Phi\otimes{\bf 1}_m$ is also CP, which follows from the tensor product property of CP map.

\subsection{Unitary-universe dilation model}\label{sec:uu}

Kraus studied the physical essence of CP map, and established the operator-sum representation (OSR) \cite{Kraus}, which forms a physical dilation model for CP map. The model we study below is termed as {\em unitary-universe dilation model} (u-u), in which a system $S$ is distinct from the rest of the universe $E$, and the universe evolves unitarily. In the usual case, suppose the initial universe state is the product of system state $\rho_S$ and environment state $\rho_E$, and the global evolution is unitary $U$, then the final state of the system, denoted as $\rho_S^t$, after tracing out environment is
\begin{align}\label{eq:kraus}\nonumber
\rho_S^t&\equiv\Phi(\rho_S)=\textrm{tr}_E(U\rho_S\otimes\rho_EU^{\dagger}) \\
&= \sum_{i,j} \lambda_i \langle e^j|U|e^i\rangle\rho_S \langle e^i|U^{\dagger}|e^j\rangle,
\end{align}
where $\rho_E=\sum_i \lambda_i|e^i\rangle\langle e^i|$, and the Kraus operator will take the form $K_{ij}=\sqrt{\lambda_i}\langle e^j|U|e^i\rangle$.
The minimal dilation is attained when the Kraus rank is $n^2$ for $n$ dimensional system.

The OSR form has the properties as follows: i) the Kraus operators are independent of system state; ii) the map $\Phi\otimes{\bf 1}_m$ acting on $\rho_{SM}$ could be written in OSR form, with $M$ as arbitrary ancilla acted on by ${\bf 1}_m$ with dimension $m$; iii) there are unitary freedom for the Kraus operators. All the properties are consistent with the requirement of the dilation theorem, which is that there cannot be initial correlation between system and environment, since the state $\rho_{SM}$ needs to be arbitrary. In other words, the requirement of CP is in conflict with the initial correlation between system and environment. This point agrees well with Pechukas's argument \cite{cpprob} (on $3$ed page) that it is not always possible to prepare a global state $\rho_{SEM}$, where system $S$ and ancilla $M$ are correlated while the reduced state of system and environment $\rho_{SE}$ is as arbitrary as we choose.

However, in physical reality, initial correlation between system and environment is possible.
This is the main origin of the CP problem. The initial condition for u-u model is $S\infty E$, and $S\nleftrightarrow E$, as shown in the left panel of Fig. \ref{fig:uu}. The dimensions of $S$ and $E$ are arbitrary.

There have been some results dealing with the initial correlation problem \cite{Sudarshan08,Lidar09,Buzek01,Sudarshan04,Carteret08,RR10}.
It is proved that a map with initial vanishing quantum discord type correlation can be written as CP map \cite{Lidar09}.
However, the map will depend on the system state, and the system state can only be in a restricted class, thus conflicting with the requirement of CP.
A positive map with initial correlation can be written in affine map form \cite{Carteret08}, yet it is not CP anymore.
One approach is developed employing one so-called assignment map ~\cite{RR10}, which needs to be linear, consistent, and positive at the same time. However, it is shown that the assignment map can only be linear at the expense of giving up positivity or the consistency.
Despite all the difficulties, it turns out by a simple {\em swap} trick studied below, which also has been used in other studies \cite{NC,nielsen97swap,guo04swap,qutritswap,genswap}, the u-u model corresponds to a CP map on the system regardless of the initial condition.

Suppose the state of system $S$ acts on $n$ dimensional Hilbert space $H$, and there exists a copy of this space which then forms a part of the environment Hilbert space, the corresponding system is denoted as ancillary-environment $A$.
Let the state of $S+E$ be $\rho_{SE}$, the state of $A$ be $\rho_A=\textrm{tr}_E(\rho_{SE})$, and the swap operator $V_{SA}$ act between the system and ancillary-environment before the evolution of the universe, and thus
\begin{align}
\rho_{SE}\otimes\rho_A=(V_{SA}\otimes{\bf 1}_E)\sigma_S\otimes\sigma_{EA}(V_{SA}^{\dagger}\otimes{\bf 1}_E),
\end{align}
where $\sigma_S=\rho_A=\textrm{tr}_E(\rho_{SE})$, $\sigma_{EA}=\rho_{SE}$, and $V_{SA}^{\dagger}V_{SA}={\bf 1}$.
As the result, the dynamics with initial condition $\rho_{SE}\otimes\rho_A$ and evolution $U_{SE}\otimes{\bf 1}_A$ is equivalent to that with initial condition $\sigma_S\otimes\sigma_{EA}$ (i.e. $\textrm{tr}_E(\rho_{SE})\otimes\rho_{SE}$) and evolution $(U_{SE}\otimes{\bf 1}_A)(V_{SA}\otimes{\bf 1}_E)\equiv W$. The swapped u-u model is depicted as the right panel in Fig. \ref{fig:uu}. The final system state $\rho_S^t$ could be expressed as
\begin{align}\label{eq:swapcp}\nonumber
\rho_S^t&=\textrm{tr}_{EA}(W\sigma_S\otimes\sigma_{EA}W^{\dagger}) \\ \nonumber
&= \sum_i p_i \textrm{tr}_{EA}(W\sigma_S\otimes|\psi_i\rangle \langle \psi_i|W^{\dagger}) \\
&= \sum_{i,j} K_{ij}\sigma_S K_{ij}^{\dagger},
\end{align}
where Kraus operator $K_{ij}=\sqrt{p_i}\langle \psi_j|W|\psi_i\rangle$, and eigenvalue decomposition $\sigma_{EA}=\sum_i p_i|\psi_i\rangle \langle \psi_i|$.
The form~(\ref{eq:swapcp}) defines a CP map which maps the system state $\sigma_S=\textrm{tr}_E(\rho_{SE})$ into $\rho_S^t$ by the set of Kraus operators $\{K_{ij}\}$. Note in the above the ancillary-environment $A$ is different from the ancilla $M$ in the discussion of CP map.
The choice of the initial state of the ancillary-environment $A$ is unique since for the case with initial correlation, given the state $\rho_{SE}$, the only reasonable initial system state is $\textrm{tr}_E(\rho_{SE})$, which is just encoded in the state of the ancillary-environment.

As to the dimension issue of the swapped u-u model, it is clear to observe that if the total dimension of the Hilbert space for $E+A$ is $n^2$, the model becomes rightly the minimal dilation model for CP map. In this case, the dimension of the Hilbert spaces for $E$ and $A$ is both $n$. In other words, we find that the swapped u-u model is indeed a particular kind of model for dilation, with bipartite environment and a special form of unitary evolution $W$. This restriction of dilation results from the initial correlation, since the formation of the initial correlation itself already reveals some information of the dynamics.

We conclude that a CP map requires dilated unitary evolution and null initial correlation, and can always be characterized by the unitary-universe dilation model, regardless of the physical initial correlation condition. As the result, we extend the validity of CP map to the case with initial correlations provided the whole universe in the model evolves unitarily.

\subsection{Convert non-completely positive map to completely positive map}\label{sec:exam}

We analyze the form of swap trick in this subsection. If the system is a single qubit, the required swap operation is simply the two-qubit swap gate $\texttt{SWAP}=\frac{1}{2}\sum_{i=0}^3\sigma_i\otimes\sigma_i$. For a qutrit system, it is the two-qutrit ternary swap gate \cite{qutritswap}. For many-body open-system with initial correlation, the dynamics is more complicated generally, since there could be multipartite correlations and $k$-local interactions. A form for general qudit swap gate and the algorithm to design it has been constructed \cite{genswap}. For practical applications, the circuit design and implementation of the swap operation are feasible.

Next, we discuss one simple yet precise example to show that how to convert non-CP map to CP map. Suppose a system qubit initially forms the maximally entangled state with the environment qubit as $|\psi\rangle=\frac{\sqrt{2}}{2}(|00\rangle+|11\rangle)$, where the states $|0,1\rangle$ can be polarization, electron, or phonon states etc in practice. The simple evolution on the two-qubit ``universe'' is assumed to be $U=e^{-i\theta X\otimes X}$, with $\theta\equiv g t$, $g$ as coupling constant, $t$ as time. The dynamics of the single-qubit system cannot be written as CP channel generally due to initial entanglement. With the method of swap trick above, one ancillary qubit is added to the universe with maximally mixed state $\rho_A=\frac{1}{2}\textrm{diag}(1,1)$. According to formula (\ref{eq:swapcp}), and using the Bell's basis to carry out the trace, the CP map on the system qubit with state $\sigma_S=\frac{1}{2}\textrm{diag}(1,1)$ will be represented by four Kraus operators corresponding to a minimal dilation,
\begin{align} \nonumber
K_1&=\frac{1}{2}e^{-i\theta}{\bf 1},\;\; K_2=\frac{1}{2}e^{-i\theta}X, \\
K_3&=\frac{i}{2}e^{-i\theta}Y,\;\; K_4=\frac{1}{2}e^{-i\theta}Z,
\end{align}
which is a single-qubit depolarization channel.
Another way to demonstrate the CP map is to show the positivity of the Choi's matrix \cite{Choi}. According to channel-state duality, a channel $\Phi$ on a qudit is CP iff the Choi's matrix $\mathcal{C}=d(\Phi\otimes {\bf 1})(|\xi\rangle\langle\xi|)$ is positive, with maximally entangled state $|\xi\rangle=\sum_{i=1}^d|ii\rangle/\sqrt{d}$. It is direct to check that the Choi's matrix for the depolarization channel in this example is the identity operator, which is clearly positive.

\subsection{Limitation of the unitary-universe model}\label{sec:pp}

The only physical requirement of quantum dynamics is positivity, and so is the tensor product of quantum dynamics (in Schr\"odinger picture). Yet, there exists mismatch between the positive map and physically realizable processes. The domain of a positive map is the whole set of state space $\mathcal{T}(H)$. While some physical operations, e.g., partial transpose, can only be realized on a restricted class of states, although they are not positive. All positive map is supposed to be realized physically. However, the problem is whether all physical dynamics is positive or not. Let us define a new class of map, the {\em physically positive} (PP) map. Observe that a physical dynamics may only live in a subspace of the whole Hilbert space and realize a restricted class of state. We can introduce the class of ``non-universal positive'' map, which acts on state $\rho$ in a subspace of $\mathcal{T}(H)$. Then the PP map contains the positive map acting on all state in $\mathcal{T}(H)$ and the`` non-universal positive'' map on a restricted class of state. As the result, we have the set relation $$H\supset PP\supset P\supset CP \supset M,$$ where we have included the set of hermitian map $H$ and Markovian dynamics $M$ for completeness. Consequently, we can describe partial transpose as a PP map acting on the class of separable state plus states which also stay positive under partial transpose.

Now let us analyze the limitations of CP map and the u-u model.
In the u-u model, the unitary evolution $U$ of the universe can be reduced to a map on system $S$ as $\Phi$, or a corresponding map on environment $E$ as $\Phi'$.
The map $\Phi$ which acts on system is CP iff $\Phi\otimes{\bf 1}_m$ is positive for any $m$ and ancilla $M$.
However, the physical problem is that the ancilla $M$ needs to exist, which cannot be a part from the environment $E$, since the unitary operator $U$ acts containing the whole environment. Thus the ancilla $M$ can only be outside of this original universe of $S+E$. This point indicates that the essence of CP map is not whether there is initial correlation or not, but instead, it is that the universe needs to evolve unitarily. For the case with initial correlation, we extend the universe with the ancillary-environment $A$, and the new universe $S+E+A$ still evolves unitarily such that ancilla $M$ is outside. As the result, for quantum dynamics beyond CP, the universe has to evolve non-unitarily. Note that if the evolution of the universe is a CP channel, then it can be further dilated to a unitary evolution resulting a CP map on system again. For the non-unitary universe model, the map on the universe is restricted as $\Phi_U\in PP \setminus CP$, resulting in another map $\Phi_S\in PP \setminus CP$ on the system. However, there will not be a precise universal formula for PP map.

\begin{figure}
\centering
\includegraphics[scale=0.28]{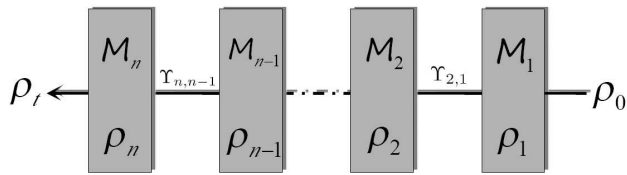}
\caption{Schematic diagram for the quantum measurement-chain model. The quantum system is prepared in $\rho_0$, the measurements $\mathcal{M}_n$ are performed on the system resulting in states $\rho_n$, $n=1,2,\cdots$, the quantum transfer matrix is $\Upsilon$. The arrow represents the direction of evolution.}
\label{fig:mchain}
\end{figure}

\section{Quantum measurement-chain formula}\label{sec:contex}

In this section, we investigate a theoretical framework different from the CP map.
For comparison, the Bloch coherence vector and affine map form is discussed in appendix \ref{sec:heis}, where its limitations are analyzed.
There are basic differences among the various frameworks to be highlighted.
The affine map form takes the quantum state as a vector, which can be visualized in the configuration space.
However, the coherence vector is generally multi-dimensional, and the dynamics of many observables need to be determined in Heisenberg picture.
While in CP map form, the quantum state is treated as a whole, i.e. one operator, so the detailed structure (e.g. different decompositions) of the state cannot be revealed.
As we know, although quantum state is a positive operator, the same density operator could be physically prepared differently. In other words, the fine-grained structure of quantum state has physical consequences, which manifests the contextual property~\cite{Spekkens}. By analogue with classical probability theory, below we develop a different technique to describe quantum dynamics, and the property of contextuality can be directly captured.

Let us recall some basic classical probability theory \cite{Breuer}. The classical system is represented as a random variable $X$, which at each particular time $t$ can take values within a set. A single-step (one-shot) stochastic dynamics is described by {\em transfer matrix} $T$. A stochastic dynamics is Markovian iff $T$ is independent of variable $X$, and $X_n$ only depends on $X_{n-1}$ and $T$. Thus, $X_n=T^nX_0$. Classical Markovian chain satisfies Chapman-Kolmogorov equation
\begin{equation}\label{eq:ck}
p(x_3,t_3|x_1,t_1)=\int dx_2 p(x_3,t_3|x_2,t_2)p(x_2,t_2|x_1,t_1).
\end{equation}
This equation has an alternative form, $p(x_3,t_3)=\int dx_2 dx_1 p(x_3,t_3|x_2,t_2)p(x_2,t_2|x_1,t_1)$.

Quantum dynamics is expressed as matrix equations (for finite dimensional case), and thus looks similar with stochastic dynamics.
However, there are differences between quantum dynamics and classical stochastic dynamics.
First, for the notion of probability, classically it originates from the ignorance of the detailed underlying causal dynamics, which means that, the uncertainty will be eliminated if the details are available. While probability in quantum dynamics comes from measurement and the related decoherence of superposition or entanglement. Furthermore, based on OSR form, quantum measurement is actually not random, that is, in a measurement $\{M_i\}$ ($\sum_i M_i^{\dagger}M_i\leq{\bf 1}$), each operation $M_i$ results in exactly one corresponding result. The other difference is that quantum dynamics contains coherence, and it is contextual, i.e., different basis can exist in reality leading to various decompositions; on the contrary, classically only one basis exists and thus the dynamics is {\em decohered} or {\em de-contexted}, for instance, the state $X$ and transfer matrix $T$ have a common basis, namely, $\{0,1,2,\dots\}$. In appendix \ref{sec:heis}, we show that the channel matrix $\mathcal{R}$ for a dynamics and the quantum state can be expressed in the same operator basis; however, quantum dynamics is one affine map instead of a linear map, and the channel matrix $\mathcal{R}$ is not a probability matrix.

\subsection{Quantum measurement transfer matrix}\label{sec:transfmatr}

A quantum measurement-chain is a chain of measurements acting on the state $\rho$ of a quantum system.
Denote the chain with $n$ measurements as $\mathcal{M}_{1\rightarrow n}=\mathcal{M}_n\circ \mathcal{M}_{n-1}\circ \cdots \circ \mathcal{M}_1$, which is depicted in Fig. \ref{fig:mchain}. A quantum dynamics can be modeled as a measurement-chain, since any two quantum states can be connected via the OSR form \cite{tong04}. The measurements do not need to be of the same rank, i.e. the number of operations. The effect of the measurement-chain is just the composition of their corresponding channels. Although state $\rho_n$ after the $n$-th measurement only depends on the measurement $\mathcal{M}_n$ and $\rho_{n-1}$, the measurement $\mathcal{M}_n$ can contain the information of the states of former steps and the initial state. Thus the measurement-chain is generally non-Markovian.
A standard Markovian dynamics is described by Lindblad equation, which can be derived from the CP map \cite{Lindblad,GKS,Alicki} with state-independent and time-independent Kraus operators. We say a quantum dynamics is non-Markovian if it can not be described by Lindblad equation and equivalently other forms such as Redfield equation.
A measurement-chain becomes Markovian if each measurement in the chain is independent of state and occupies infinitesimal time; then the generator of the channel is the sum of those for the building channels via Lie-Trotter decomposition \cite{Tro59}.
In addition, there are different quantum Markovian-chain theories in literatures, e.g. quantum walk theory \cite{ADZ}, and consistent history theory \cite{history}, yet there is no measurement-chain.

Let us construct a measurement-chain with two measurements $\mathcal{M}=\{M_i\}$ and $\mathcal{N}=\{N_j\}$ for the sake of convenience, and further generalization is straightforward. Note the measurements could depend on system state generally. Suppose there is no evolution except during the measurement processes. A quantum state prepared in $\rho_0$, after the first measurement $\mathcal{M}$, becomes
\begin{equation}\label{}
\rho_1=\sum_i M_i \rho_0M_i^{\dagger},
\end{equation}
which can also be written as $\rho_1=\sum_i p_i \sigma_i$, with $p_i=\textrm{tr}(M_i^{\dagger}M_i\rho_0)$, $\sum_i p_i=1$. After the second measurement $\mathcal{N}$, the state becomes
\begin{equation}\label{}
\rho_2=\sum_j N_j\rho_1 N_j^{\dagger},
\end{equation}
with $q_j=\textrm{tr}(N_j^{\dagger}N_j\rho_1)$, $\sum_j q_j=1$. The probability for the second measurement can be decomposed as
\begin{eqnarray}\nonumber
q_j&=& \textrm{tr}(N_j^{\dagger}N_j\rho_1) \\ \nonumber
&=& \sum_i \textrm{tr}(N_j^{\dagger}N_jM_i \rho_0M_i^{\dagger})\\ \nonumber
&=& \sum_i \frac{\textrm{tr}(M_i^{\dagger}N_j^{\dagger}N_jM_i \rho_0)}{\textrm{tr}(M_i^{\dagger}M_i\rho_0)}\textrm{tr}(M_i^{\dagger}M_i\rho_0)\\
&\equiv& \sum_i m_{ji}p_i,
\end{eqnarray}
$m_{ji}$ forms the {\em quantum measurement transfer matrix} $\Upsilon$, which is the analogue of classical transfer matrix $T$. The quantum states after each measurement are represented via the {\em probability vector} $\vec{Q}=(q_1,q_2,\cdots)$, and $\vec{P}=(p_1,p_2,\cdots)$, and therefore
\begin{align}
\vec{Q}=\Upsilon \vec{P}.
\end{align}
A general form is expressed as $\vec{P}_n=\Upsilon_{n,n-1}\vec{P}_{n-1}$, where $\Upsilon_{n,n-1}$ transfers vector $\vec{P}_{n-1}$ for state $\rho_{n-1}$ to $\vec{P}_n$ for state $\rho_n$ ($n\geq2$). $m_{ji}$ could be called the {\em quantum transition probability}, and has the properties
\begin{equation}\label{}
\sum_j m_{ji}=1, \sum_{i,j}p_i m_{ji}=1.
\end{equation}
$\Upsilon$ is a lefty stochastic matrix, of which each column sums up to one and all elements are nonnegative. If the ranks of the two measurements are the same, $\Upsilon$ is square matrix. For Markovian case, by comparison with Chapman-Kolmogorov equation (\ref{eq:ck}), the correspondence is $q_j\sim p(x_3,t_3)$, $p_i\sim p(x_2,t_2|x_1,t_1)$, $m_{ji}\sim p(x_3,t_3|x_2,t_2)$.

Furthermore, the matrix $\Upsilon$ has the feature that its formula depends on the measurement steps and the quantum state.
As we know the channel formula is fixed for a simple Markovian quantum dynamics, however, the matrix $\Upsilon$ varies generally, which we
treat as the manifestation of the property of contextuality.
Open-system quantum dynamics is contextual generally.
A context is formed by arbitrary two successive measurements $\mathcal{M}_{n-1}$, $\mathcal{M}_n$, and the quantum state $\rho_{n-2}$ on which the measurement $\mathcal{M}_{n-1}$ is implemented. Notably, we find that when the measurements in the chain reduce to projective measurements, the matrix $\Upsilon$ will not depend on the system state, and thus the dynamics is ``de-contexted'', and then we say there is no contextuality anymore. For the construction above of the measurement-chain with two measurements, if they are projection-valued measurement (PVM), and binary for simplicity, namely, $\mathcal{M}=\{|m^+\rangle\langle m^+|,|m^-\rangle\langle m^-|\}$ and $\mathcal{N}=\{|n^+\rangle\langle n^+|,|n^-\rangle\langle n^-|\}$, the matrix $\Upsilon$ takes the form
\begin{align}\label{eq:pvm}
\Upsilon=\left(
  \begin{array}{cc}
    |\langle m^+|n^+\rangle|^2 & |\langle m^-|n^+\rangle|^2 \\
    |\langle m^+|n^-\rangle|^2 & |\langle m^-|n^-\rangle|^2 \\
  \end{array}
\right),
\end{align}
which is a doubly stochastic and symmetric matrix. If there exists evolution between the measurement processes, it will not be symmetric. It is interesting that the measurement transfer matrix formula relates to the scattering matrix (S-matrix) approach to particle physics, and the general form of $\Upsilon$ with POVM measurements could be treated as a generalization of the S-matrix. For the chain of PVM, if each measurement is the same, the matrix $\Upsilon$ becomes identity operator.

The property of contextuality of quantum dynamics concerns the differences between PVM and POVM. In the chain of PVM, the information of the initial quantum state is lost, so in this sense, PVM is maximally destructive. On the contrary, POVM does not destroy the information
of the initial quantum state completely at once, so there will be coherence and thus contextuality in the dynamics, until the system is eventually decohered. We will analyze some examples in the next section to demonstrate the points.

\subsection{Examples: contextuality in quantum dynamics}\label{sec:excont}

In this section we employ the quantum measurement-chain formula to analyze some standard physical models. First, let us consider the exponential decay process described as amplitude-damping channel. Starting from the state $|\psi_0\rangle=\alpha|g\rangle+\beta|e\rangle$, the system decays to the ground state $|g\rangle$ for equilibrium. We model the decay process as a chain of POVM measurements, each of which takes time $\tau$, and the two Kraus operators are $K_1=\begin{pmatrix} 1 & 0 \\ 0 & \sqrt{\gamma} \end{pmatrix}$, $K_2=\begin{pmatrix} 0 & \sqrt{1-\gamma} \\ 0 & 0 \end{pmatrix}$, where $\gamma=e^{-\tau/T}$, $T$ is the life time of excited state $|e\rangle$ \cite{NC}. The system state after the $n$-th measurement is
\begin{align}
\rho_n=\begin{pmatrix}
  1-\beta^2\gamma^n & \alpha\beta\gamma^{n/2} \\
  \alpha\beta\gamma^{n/2} & \beta^2\gamma^n
\end{pmatrix}.
\end{align}
The measurement transfer matrix between the $(n-1)$-th and $n$-th measurements takes the form
\begin{align}
\Upsilon_{n,n-1}=\begin{pmatrix}
  \frac{1-\gamma^{n-2}+\gamma^n}{1-\gamma^{n-2}+\gamma^{n-1}} & 1 \\
  \frac{\gamma^{n-1}-\gamma^n}{1-\gamma^{n-2}+\gamma^{n-1}} & 0
\end{pmatrix},
\end{align}
and the asymptotic form at equilibrium is $\Upsilon_{\infty}=\begin{pmatrix} 1 & 1 \\ 0 & 0 \end{pmatrix}$. The relations $\Upsilon_{n,n-1}^{21}=\textrm{tr}(K_2^{\dagger}K_2K_1\rho_{n-2}K_1^{\dagger})/\textrm{tr}(K_1^{\dagger}K_1\rho_{n-2})$ etc are used to calculate the elements of $\Upsilon_{n,n-1}$. It is clear that the formula of the measurement transfer matrix $\Upsilon_{n,n-1}$ depends on the measurement steps. From its formula, we can see that the transition probability from $K_2$ to $K_1$ is $1$, which intuitively manifests the roles of the two Kraus operators, namely, $K_2$ causes the decay of the excited state population, while $K_1$ causes both decay and decoherence, and in the long-time limit the effect of $K_1$ dominates. Although decay process also exists classically and some quantum decay processes can be modeled by classical dynamics, however, classical decay process is not contextual, since no coherence exists at least classically. As the result, we see that the quantum decay process is contextual, and contextuality captures the primary properties of quantum dynamics.

Next, we analyze the Stern-Gerlach experiment with sequential measurements, where no contextuality exists.
A beam of electrons will split into two branches after passing through the magnetic field.
The effect of the magnetic field is a selective quantum measurement, i.e. the branches of the state are selectively distinguished via different spatial paths. Note the POVM in quantum decay process above is not selective, though.
Suppose the sequential measurements are alternate along $z$-axis and $x$-axis. Based on formula (\ref{eq:pvm}), the transfer matrix becomes
\begin{align}
\Upsilon=\begin{pmatrix}
  \frac{1}{2} & \frac{1}{2} \\
  \frac{1}{2} & \frac{1}{2}
\end{pmatrix},
\end{align}
which is all the same for each step, and thus the dynamics is de-contexted.
Still, there exists quantum property responsible for the Stern-Gerlach experiment, which is the superposition of quantum state. The state $|\uparrow_x\rangle$ will be a superposed state when expanded and measured in the basis of $|\uparrow_z,\downarrow_z\rangle$. This example indicates that the notion of contextuality is different from superposition, yet both of them rely on the existence of quantum coherence.

\subsection{Weak-measurement transfer matrix}\label{sec:wtm}

In this section, we show that the formula of measurement transfer matrix can be generalized. Suppose the first measurement $\mathcal{M}$ on system $\rho_0$ is weak, in the sense that we record some information of the system (observable) without changing the state significantly \cite{aav}, the system state becomes $\tilde{\rho}_0\simeq\rho_0$, and then perform the second measurement $\mathcal{N}$ which provides the post-selection. The probability becomes
\begin{eqnarray}\label{eq:wprob}\nonumber
\tilde{q}_j&=& \textrm{tr}(N_j^{\dagger}N_j\tilde{\rho}_0) \\ \nonumber
&=& \sum_i \textrm{tr}(M_i^{\dagger}M_iN_j^{\dagger}N_j\tilde{\rho}_0)\\ \nonumber
&=& \sum_i \frac{\textrm{tr}(M_i^{\dagger}M_iN_j^{\dagger}N_j\tilde{\rho}_0)}{\textrm{tr}(M_i^{\dagger}M_i\tilde{\rho}_0)}\textrm{tr}(M_i^{\dagger}M_i\tilde{\rho}_0)\\
&\equiv& \sum_i \tilde{m}_{ji}\tilde{p}_i,
\end{eqnarray}
$\tilde{m}_{ji}$ forms the {\em weak-measurement transfer matrix} $\tilde{\Upsilon}$. It has the properties
\begin{equation}
\sum_j \tilde{m}_{ji}=1, \sum_{i,j}\tilde{p}_i \tilde{m}_{ji}=1.
\end{equation}
$\tilde{\Upsilon}$ is a generalized lefty stochastic matrix, not all elements are real. It relates to {\em weak value} \cite{aav}, a generalization of eigenvalue and expectation value, which even can take complex values. Suppose the two measurements are performed to measure observable $A$. Then, in weak measurement, with probability $\tilde{q}_j$ we measure the weak value $\varpi_A^j$ which takes the form
\begin{align}\label{}
\varpi_A^j=\frac{\textrm{tr}(N_j^{\dagger}N_jA\tilde{\rho}_0)}{\textrm{tr}(N_j^{\dagger}N_j\tilde{\rho}_0)},
\end{align}
with $\langle A\rangle= \textrm{tr}(A\tilde{\rho}_0)=\sum_j \varpi_A^j \tilde{q}_j$. The weak value $\varpi_A^j$ reduces to the original form for PVM case \cite{aav}. Note that in equation (\ref{eq:wprob}), $N_j^{\dagger}N_j$ can also be viewed as one observable with expectation value $\tilde{q}_j$, then the weak transition probability $\tilde{m}_{ji}$ is just the weak value of $N_j^{\dagger}N_j$ pre-conditioned on
the operation $M_i^{\dagger}M_i$.

\section{Discussion and Conclusion}\label{sec:conc}

In this work, the properties of open-system quantum dynamics, namely positivity and contextuality, were studied.
These two notions are defined from different standpoints (such as there are differences between nonlocality and entanglement).
Quantum dynamics is distinctive from classical mechanics and stochastic dynamics, but there is no unique concept that could capture the essence of quantumness or openness completely. From our study we can see that the property of contextuality is much more basic than complete positivity.
We demonstrated the limitation of completely positive (CP) map. We verified that the CP map can be modeled physically by the unitary-universe dilation model, which provides an elegant method to resolve the initial correlation problem. As a subset of physically positive map and hermitian map, CP map is not sufficient to describe all kinds of quantum dynamics. Therefore we develop the quantum measurement-chain formula, which is natural to describe the property of contextuality of quantum dynamics. The fine-grained structures of quantum operators (including density matrix, observable, and measurement etc) would result in realistic effects since quantum dynamics occurs not only in Hilbert space, but also in spacetime associated with quantum reference frame such that different basis and decompositions exist physically. Quantum contextuality directly relies on the fine-grained structure of quantum operator and quantum coherence. The traditional way to reveal contextuality is to employ inequalities of expectation values of non-commuting observables, which indeed relies on the fine-grained structure of operators too. The formula of measurement transfer matrix can be viewed as a new way to understand contextuality.

The framework of using swap operation to resolve the initial correlation problem could have further applications in channel theory, and quantum control etc. Also, problems of quantum and semi-classical non-Markovian quantum dynamics with initial correlation could be studied by applying swap trick. The measurement, and weak-measurement transfer matrix formula can be used in measurement theory, and analysis of quantum-classical transition etc.

\section*{Acknowledgements}
This work was supported by the National Basic Research Program of China (Grant No. 2009CB929400).
DSW thank C\'{e}sar A Rodr\'{i}guez-Rosario for helpful discussions.

\appendix

\section{Heisenberg picture}\label{sec:heis}

An affine map form associated with Bloch coherence vector, which is geometric and more general than CP map, can describe any hermitian (not necessarily positive) dynamics. In generalized Bloch ball, the quantum state is represented as a coherence vector, whose evolution manifests the quantum dynamics. This geometric picture is mostly analogue of classical mechanics, namely, the dynamics of classical position or velocity vector can also be expressed as affine map. The Hilbert space equipped with Hilbert-Schmidt operator inner product $\langle A,B\rangle=\textrm{tr}(A^{\dagger} B)$ is suitable to describe open-system quantum dynamics. For a qudit system, there exists a complete orthonormal operator basis $\{F_i, i=0,2,\dots,d^2-1\}$, and $\textrm{tr}(F_i^{\dagger}F_j)=\delta_{ij}$, with $F_0=\frac{1}{\sqrt{d}}{{\bf 1}}$, the others traceless. The qudit system can be decomposed as $\rho=({\bf 1}+\sum_{i=1}^{d^2-1}f_iF_i)/d$, with $f_i$ as the expectation value of observable $F_i$. The Bloch coherence vector is formed as a column vector $\vec{F}=(f_1,f_2,\cdots,f_{d^2-1})$. The open-system dynamics is then represented by the Bloch vector, i.e. the evolution of the $d^2-1$ parameters $f_i$. Heisenberg picture has one merit that there exists one well-defined form for any hermitian map \cite{hermi} on a system observable $\mathcal{O}_S$, which is
\begin{align}\label{eq:hp}
\Phi(\mathcal{O}_S)=\sum_{\alpha,\beta} d_{\alpha\beta} K^{\dagger}_{\alpha}\mathcal{O}_SK_{\beta},
\end{align}
where $d_{\alpha\beta}$ forms a hermitian matrix. Once we know the form of the map, the dynamics for each $F_i$ then the expectation value $f_i$ can be derived. Another equivalent way is to represent the map in the basis of $F_i$ as a matrix $\mathcal{R}$, with elements $\mathcal{R}_{ij}=\textrm{tr}(F_i\Phi(F_j))/d$.
Note, the $\mathcal{R}$ matrix will relate to Choi's matrix $\mathcal{C}$ in one elegant way $\langle ik|\mathcal{R}|jl\rangle=\langle ij|\mathcal{C}|kl\rangle$, via involution. The channel matrix $\mathcal{R}$ takes the form \cite{King2001}
\begin{align}\label{}
\mathcal{R}=\begin{pmatrix}
1 & \vec{0} \\
\vec{r} & R
\end{pmatrix},
\end{align}
and the dynamics becomes affine map $\vec{F}'=R\vec{F}+\vec{r}$, and $\rho'=\mathcal{R}\rho$.

It seems it is better to employ the Bloch vector dynamics via hermitian map or affine map formulas in Heisenberg picture than the positive map formula in Schr\"odinger picture. The Bloch vector dynamics is the quantum analogue of classical mechanics, however, the geometrical visualization is only proper for low dimensional case, e.g. qubit dynamics in NMR research.

Below, we present the CP map form on system observable $\mathcal{O}_S$. For the u-u model, one environment is needed with the environment observable ${\bf 1}_E$. Note for the initial correlated case, the environment is bipartite (including the ancilla). Denote the global unitary as $U$. In order to find the OSR form for the system observable, we apply the fact that
\begin{align}\nonumber
\langle \mathcal{O}_S \rangle &= \textrm{tr}_S(\mathcal{O}_S\rho_S^t)=\textrm{tr}_S(\mathcal{O}_S^t\rho_S) \\ \nonumber
&= \textrm{tr}_{SE}((U^{\dagger}\mathcal{O}_S\otimes{\bf 1}_E U)\rho_S\otimes \rho_E) \\ \nonumber
&= \textrm{tr}_S (\sum_{ij}  \lambda_i \langle e^j| (U^{\dagger}\mathcal{O}_S\otimes{\bf 1}_E U)\rho_S |e^i\rangle\langle e^i|e^j\rangle) \\
&= \textrm{tr}_S (\sum_i  \lambda_i \langle e^i| U^{\dagger}\mathcal{O}_S\otimes{\bf 1}_E U |e^i\rangle \rho_S),
\end{align}
then the OSR form of the system observable $\mathcal{O}^t_S$ is
\begin{align}\label{eq:kraush}\nonumber
\mathcal{O}_S^t &= \sum_i  \lambda_i \langle e^i| U^{\dagger}\mathcal{O}_S\otimes{\bf 1}_E U |e^i\rangle \\ \nonumber
&= \sum_{ij}  \lambda_i \langle e^i| U^{\dagger}\mathcal{O}_S\otimes|e^j\rangle\langle e^j| U |e^i\rangle \\
&= \sum_{i,j} K_{ij}^{\dagger}\mathcal{O}_S K_{ij},
\end{align}
with Kraus operator $K_{ij}=\sqrt{\lambda_i}\langle e^j| U |e^i\rangle$, consistent with forms (\ref{eq:kraus}). The above formula applies to both cases (with/without initial correlation) for the u-u model. A special form of hermitian map can be realized, where the pure state vector $|e^i\rangle$ can be further decomposed as $|e^i\rangle=\sum_{\alpha}c_{\alpha}|\alpha_i\rangle$, the Kraus operator $K_{ij}$ becomes $K_{ij}=\sqrt{\lambda_i}\sum_{\alpha,\alpha'}c_{\alpha}c^*_{\alpha'} \langle \alpha'_j|U|\alpha_i\rangle$. Then it is direct to verify that when the environment observable reduces to a projector instead of identity, and the environment state becomes pure, the CP map reduces to the form of hermitian map (\ref{eq:hp}), with separable coefficients $d_{\alpha\beta}=c_{\alpha}c_{\beta}^*$.


\begin{thebibliography}{00}
\bibitem{NC} M. A. Nielsen and I. L. Chuang, {\em Quantum Computation and Quantum Information} (Cambridge University Press, Cambridge, England, 2000).
\bibitem{Breuer} H.-P. Breuer and F. Petruccione, {\em The Theory of Open Quantum Systems} (Oxford University Press, New York, 2002).
\bibitem{Stinespring} W. F. Stinespring, Proc. Am. Math. Soc. {\bf 6} (1955) 211.
\bibitem{Choi} M.-D. Choi, Numer. Linear Algebra Appl. {\bf 10} (1975) 285.
\bibitem{Chattaraj} P. K. Chattaraj, {\em Quantum trajectories} (CRC Press, 2011).
\bibitem{Lindblad} G. Lindblad, Commun. Math. Phys. {\bf 48} (1976) 119.
\bibitem{GKS} V. Gorini, A. Kossakowski, and E. C. G. Sudarshan, J. Math. Phys. {\bf 17} (1976) 821.
\bibitem{Alicki} R. Alicki and M. Fannes, {\em Quantum Dynamical Systems} (Oxford University Press, Oxford, 2001).
\bibitem{Kraus} K. Kraus, {\em States, Effects and Operations: Fundamental Notions of Quantum Theory} (Springer, Berlin, 1983).
\bibitem{Sudarshan08} C. A. Rodr\'{i}guez-Rosario, K. Modi, A.-M. Kuah, A. Shaji, and E. C. G. Sudarshan, J. Phys. A {\bf 41} (2008) 205301.
\bibitem{Lidar09} A. Shabani and D. A. Lidar, Phys. Rev. Lett. {\bf 102} (2009) 100402.
\bibitem{tong04} D. M. Tong, L. C. Kwek, C. H. Oh, J.-L. Chen, and L. Ma, Phys. Rev. A {\bf 69} (2004) 054102; D. Salgado, J. L. S\'{a}nchez-G\'{o}mez, and M. Ferrero, Phys. Rev. A {\bf 70} (2004) 054102.
\bibitem{ks} E. P. Specker, Dialectica {\bf 14} (1960) 239; S. Kochen and E. P. Specker, J. Math. Mech. {\bf 17} (1967) 59.
\bibitem{Cabello} A. Cabello, Phys. Rev. Lett. {\bf 101} (2008) 210401.
\bibitem{measurec} A. Grudka, K. Horodecki, M. Horodecki, P. Horodecki, R. Horodecki, P. Joshi, W. K{\l}obus, and A. W\'{o}jcik, arXiv:1209.3745.
\bibitem{Spekkens} R. W. Spekkens, Phys. Rev. A {\bf 71} (2005) 052108.
\bibitem{Bhatia} R. Bhatia, {\em Positive definite matrices} (Princeton University Press, 2007).
\bibitem{cpprob} P. Pechukas, Phys. Rev. Lett. {\bf 73} (1994) 1060.
\bibitem{SMR} E. C. G. Sudarshan, P. M. Mathews, and J. Rau, Phys. Rev. {\bf 121} (1961) 920.
\bibitem{Buzek01} P. \v{S}telmachovi\v{c} and V. Bu\v{z}ek, Phys. Rev. A {\bf 64} (2001) 062106.
\bibitem{Sudarshan04} T. F. Jordan, A. Shaji, and E. C. G. Sudarshan, Phys. Rev. A {\bf 70} (2004) 052110.
\bibitem{Carteret08} H. A. Carteret, D. R. Terno, and K. \.{Z}yczkowski, Phys. Rev. A {\bf 77} (2008) 042113.
\bibitem{RR10} C. A. Rodr\'{i}guez-Rosario, K. Modi, and A. Aspuru-Guzik, Phys. Rev. A {\bf 81} (2010) 012313.
\bibitem{nielsen97swap} M. A. Nielsen and C. M. Caves, Phys. Rev. A {\bf 55} (1997) 2547.
\bibitem{guo04swap} Y.-J. Gu, C.-M. Yao, Z.-W. Zhou, and G.-C. Guo, J. Phys. A: Math. Gen. {\bf 37} (2004) 2447.
\bibitem{qutritswap} Y. M. Di, J. Zhang, and H. R. Wei, Science in China Series G {\bf 51} (2008) 1668.
\bibitem{genswap} C. M. Wilmott and P. R. Wild, Int. J. Quanum Inform. {\bf 10}, (2012) 1250034.
\bibitem{Tro59} H. Trotter, Proc. Am. Math. Soc.  {\bf 10} (1959) 545.
\bibitem{ADZ} Y. Aharonov, L. Davidovich, and N. Zagury, Phys. Rev. A {\bf 48} (1993) 1687.
\bibitem{history} R. B. Griffiths, J. Stat. Phys. {\bf 36} (1984) 219; R. Omn\'es, Rev. Mod. Phys. {\bf 64} (1992) 339.
\bibitem{aav} Y. Aharonov, D. Z. Albert, and L. Vaidman, Phys. Rev. Lett. {\bf 60} (1988) 1351.
\bibitem{hermi} R. D. Hill, Linear Algebra Appl. {\bf 6} (1973) 257.
\bibitem{King2001} C. King and M. B. Ruskai, IEEE Trans. Inf. Theory {\bf 47} (2001) 192.
\end{thebibliography}
\end{document}